# Supervised learning of an opto-magnetic neural network with ultrashort laser pulses

A. Chakravarty, J.H. Mentink*, C. S. Davies, K. Yamada, A.V. Kimel and Th. Rasing*
Radboud University, Institute for Molecules and Materials,
Heyendaalseweg 135, 6525 AJ, Nijmegen, the Netherlands
*Correspondence to: j.mentink@science.ru.nl, th.rasing@science.ru.nl

The explosive growth of data and its related energy consumption is pushing the need to develop energy-efficient brain-inspired schemes and materials for data processing and storage. Here, we demonstrate experimentally that Co/Pt films can be used as artificial synapses by manipulating their magnetization state using circularly-polarized ultrashort optical pulses at room temperature. We also show an efficient implementation of supervised perceptron learning on an opto-magnetic neural network, built from such magnetic synapses. Importantly, we demonstrate that the optimization of synaptic weights can be achieved using a global feedback mechanism, such that the learning does not rely on external storage or additional optimization schemes. These results suggest there is high potential for realizing artificial neural networks using optically-controlled magnetization in technologically relevant materials, that can learn not only fast but also energy-efficient.

The rapid growth of modern information and communication technology (ICT) has led to an enormous increase in energy consumption, which is already now around 7% of the global electrical energy production. Owing to the inherently energy-efficient brain-inspired computing principles, implementing such neuromorphic architectures offers enormous potential to dramatically reduce the energy consumption of ICT. Magnetic materials are already at the center of computing today, due to their ability to store information within the direction of magnetic moments in a non-volatile and rewritable way. In recent years, tremendous progress has been made in controlling magnetism with femtosecond optical pulses, including demonstrations of record-breaking fast write-read events[1], operation in technologically relevant materials such as Co/Pt[2] and enabling magnetic recording that is not only much faster but also exhibits a projected heat load of only 22 aJ per magnetic bit[3]. These demonstrations suggest that all-optical manipulation of magnetism offers great potential to realize neural networks that can be trained not only much faster but also much more efficiently than all-electrical material implementations under development today[4-9]. In particular, similar as for purely optical neural networks[10-12] operation proceeds ultrafast and at very low energy cost. Moreover, combining this with magnetism integrates inherent non-volatility and optical adaptability, which potentially yields fast and energy-efficient learning as well. However, so far it has not been demonstrated that it is possible to exploit optical control of magnetism for brain-inspired computing. Here, we study the control of the magnetization state of Co/Pt thin films in response to picosecond optical pulses. We experimentally demonstrate that the cumulative all-optical switching process in these materials[13, 14] provides an energy-efficient mechanism for realizing artificial synapses, in which the internal state of the synapse is stored with non-volatility and can be controlled continuously and reversibly using the helicity of light. Moreover, we realize supervised learning of a network of two such opto-magnetic synapses, that together form a single-layer perceptron. Importantly, we exploit iterative optimization of synaptic weights using a global feedback mechanism. Hence, the learning does not rely on external storage or additional optimization strategies, which simplifies the implementation.

The perceptron is an elementary model of neural computation[15]. An example of the simplest single-layer perceptron is shown in Figure 1a. It has inputs $x_1, \ldots, x_n$ that are connected to adaptable synaptic weights $w_1, \ldots, w_n$. The output $O$ is a nonlinear function of the weighted inputs $y_i = w_i x_i$ and the threshold $b$: $O = \text{sign}(\sum y_i - b)$, where we introduced the sign function as a simple nonlinear function. In this form, the perceptron can be interpreted as a neuron that fires when the sum of the weighted inputs is larger than a threshold. We focus on the simple perceptron as an example of a neural network that can learn (by adapting synaptic weights) a given function, also known as supervised learning. To this end, several input patterns $x_1^\mu, \ldots, x_n^\mu$, µ=1,2,..,p with desired outputs $O_d^\mu$ are provided. The perceptron learning rule can then be defined as an iterative procedure where one cycles over the input patterns. At each pattern the actual output $O^\mu$ is evaluated and the weights are changed according to a global error $\text{E} = \text{sign}(O_d^\mu - O^\mu)$:

$$w_i = w_i + \Delta w_i \quad \text{Eq. (1)}$$
$$\Delta w_i = \eta\, x_i^\mu\, \text{E} \quad \text{Eq. (2)}$$

Here, $\eta \ll 1$ is the learning rate. Learning stops when for all patterns the desired output is realized ($\text{E} = 0$). Note that, although a global feedback is used, weights are changed locally due to the dependence on $x_i^\mu$ in Eq. (2). Importantly, this iterative learning procedure does not necessitate additional external storage nor external optimization. Instead, it can exploit the non-volatile local storage of synaptic weights in the material itself. Implementing a perceptron in a physical system therefore requires continuous adaptable and non-volatile synaptic weights as well as multiplication of the weights with given inputs. Below we describe how this is realized using a magneto-optical setup, and subsequently show how learning is achieved.

To implement the multiplication, we exploit the concept of a polarizing microscope that is illustrated in Figure 1b. The magnetic sample that features magnetization parallel or antiparallel to the optical wave vector is placed between two polarization filters. Due to the magneto-optical Faraday effect, the axis of optical polarization is rotated clockwise (anticlockwise) via the magneto-optical Faraday effect. The relative angle between the polarizers is tuned such that minimum (maximum) light passes through the analyzer when the light transmits through magnetic domains pointing parallel (antiparallel) to the light's direction i.e. pointing up or down.

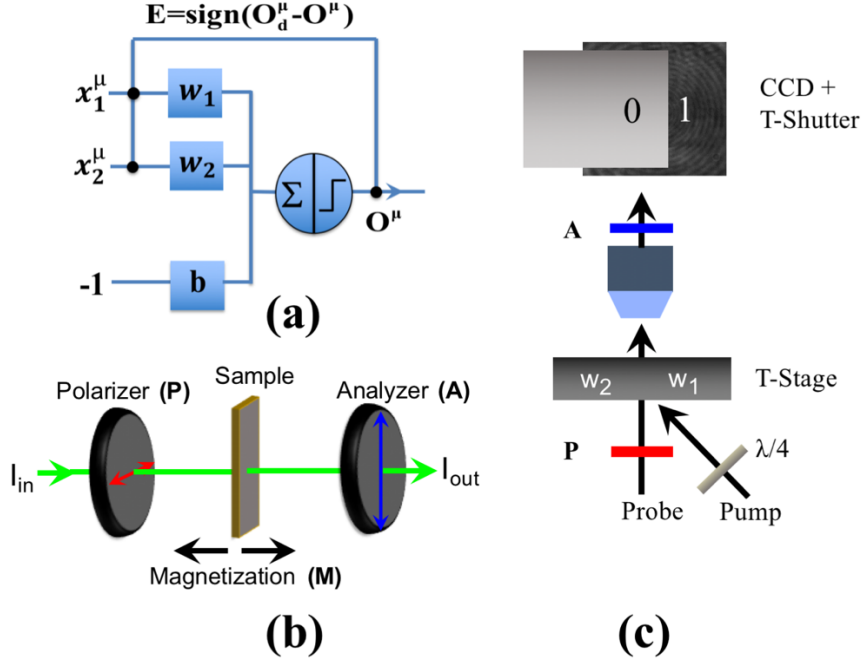

**Figure 1:** (a) Schematic illustration of the global feedback mechanism for a single-layer perceptron with adaptable synapses $w_i$, threshold b and inputs $x_i$. (b) implementation of an artificial synapse using the remnant magnetization of a magnetic medium placed between two nearly-crossed polarizers, P and A. (c) The weight is continuously adapted by optical control of the magnetization using pump pulses. Translating the sample (T-Stage) allows selecting the positions of distinct synaptic weights. Computer-controlled rotation of the quarter wave plate (λ/4) is used to determine the sign of pump-induced magnetization changes. Binary inputs are implemented by moving a shutter (T-Shutter) in front of the CCD camera.

From analyzing the light power $P \propto I_{out}$ that is locally detected by the CCD camera we obtain, for small Faraday rotations as applicable here, a linear relation with the incoming illumination $I_{in}$:

$$I_{out} = KM I_{in}. \qquad \text{Eq. (3)}$$

Here $K$ depends on the material-specific Verdet constant, the thickness $d$ and magnetization $m$ of the material, while $M = d(S_1 - S_2)m$ is the net magnetic moment, where $S_1$ ($S_2$) denotes the area of up (down) domains. By identifying $I_{out}$ with the weighted input $y_i$, '$KM$' with the synaptic weight $w_i$, and $I_{in}$ with the synaptic input $x_i$, we directly obtain the desired multiplication.

To realize adaptable and non-volatile synaptic weights, we optically control the magnetization *M*. Therefore, a magnetic material with strong out-of-plane magnetic anisotropy and a strong opto-magnetic response is needed. Moreover, instead of exploiting single-shot binary switching seen in ferrimagnetic alloys and multilayers[16-18], reversible gradual changes of magnetization in response to laser pulses are required. To this end, the multi-shot helicity-dependent control of magnetization in ferromagnetic Co/Pt thin films is an attractive candidate[2, 13, 18]. In this study, we used a sputter-grown Co/Pt multilayered thin film structure, mounted on a synthetic quartz glass substrate, with nanometer thickness composition of Glass/Ta(3)/Pt(3)/Co(0.6)/Pt(3)/MgO(2)/Ta(1). The structure has a coercive field of 20 mT.

The magneto-optical setup in which both the multiplication and optical control of synaptic weights are implemented is schematically shown in Figure 1c. Gaussian pump pulses of 800 nm central wavelength and duration of 4 ps were selected from a 1kHz repetition rate Ti:sapphire amplified laser system. The pump pulse was targeted to hit the sample at about 10° incidence angle, and focused to a spot with a Gaussian waist of 38 µm. The measured incident optical fluence was maintained at 1.3mJ/cm$^2$. A rotatable quarter wave plate was used to select left/right circularly polarized pump pulses to either decrease/increase the net magnetic moment of the sample and thereby adapt the synaptic weight. For ease of demonstration a continuous wave laser of 532 nm wavelength was used for probing.

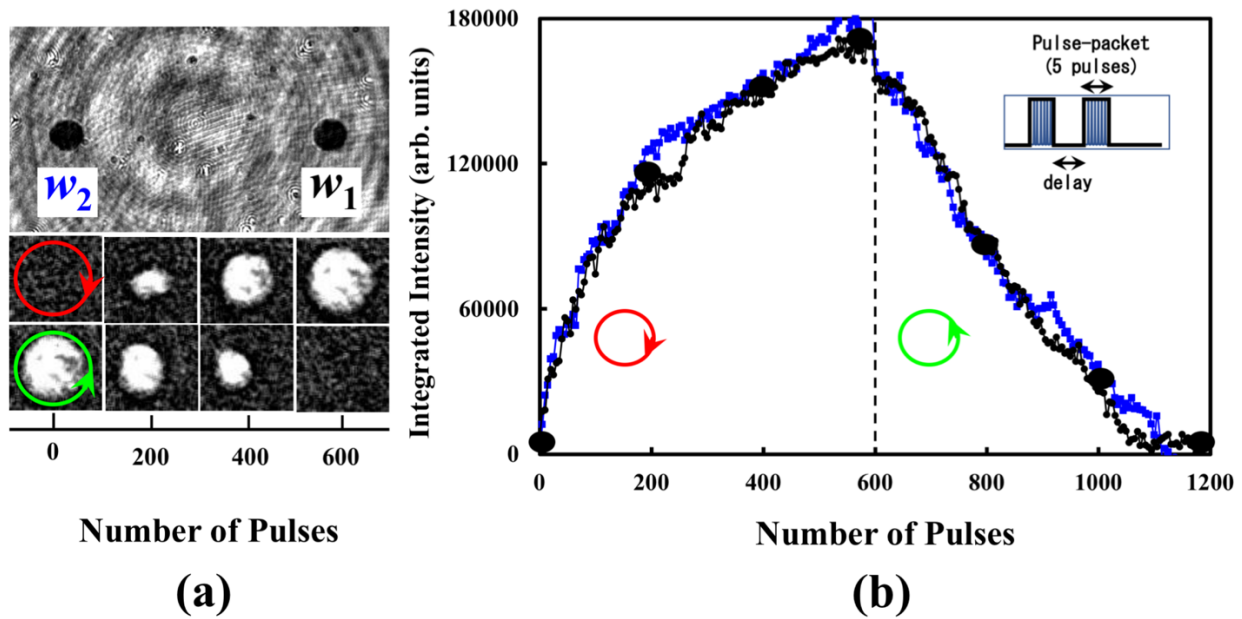

Figure 2: (a) A raw image of two physically-separated laser-written weight spots, $w_1$, $w_2$ (dark spots); the bottom shows eight background subtracted images illustrating evolution of the magnetization change due to the right (top row) and left (bottom row) circularly-polarized pump pulses irradiating the Co/Pt sample. (b) Extracted intensity changes as function of the number of pump pulses, demonstrating continuously-controllable weights, shown in black (blue) for $w_1$ ($w_2$). Black solid disks correspond to the images in the bottom rows of (a). The inset shows the pulse packets used for learning.

In Figure 2 we study the response of the magnetization in the Co/Pt thin film to laser pulses. Without laser irradiation the selected area of the sample shows a nearly uniform light gray contrast. Two positions of the sample are selected for the definition of synaptic weights $w_1, w_2$. These are seen in Figure 2a as dark regions with opposite magnetization and are obtained by irradiation with 600 right circularly polarized laser pulses, sufficient to yield saturation. The eight images at the bottom of Figure 2a show experimental results of pumping with laser pulses for right (top row) and left (bottom row) circular polarization. For clarity the background magnetization of the sample prior laser writing was subtracted. We observe that for right circularly-polarized light the intensity, and thus the switched domain area, grows with the number of pulses. Conversely, for left circularly-polarized light the previously large domain area decreases in size, as the number of pulses increases.

To quantify the adaptability, Figure 2b shows the integrated pixel intensity detected by the CCD camera as function of the number of right handed circularly polarized pump pulses for both synaptic weights. A maximum of N=600 pulses, each of 4 ps duration was used. We find that the magnetic contrast changes approximately linearly with the number of pulses. Deviation from purely linear behavior is attributed to deviations of the polarization state, that were estimated to be about 5% by comparison of initial and final ellipticity. After 600 pulses, the pump helicity was changed to left circular polarization, and a nearly linear decrease of the integrated intensity (ending close to the initial value) was observed after exposure to another 600 pulses. Hence, using the Co/Pt sample we demonstrate continuous and reversible laser-writing of nonvolatile magnetic synaptic weights in accordance with Eq. (1).

To implement supervised learning, we focused on the simplest example of a single-layer perceptron with two inputs. This artificial neural network can learn linearly separable problems[15] such as the logical AND and OR functions. For two inputs there are four input patterns ($\mu=1..4$): $(x_1^1, x_2^1)=(0,0)$; $(x_1^2, x_2^2)=(1,0)$; $(x_1^3, x_2^3)=(0,1)$; and $(x_1^4, x_2^4)=(1,1)$. These input patterns are implemented by placing a translational shutter between the magneto-optical microscope and the CCD camera. As shown in Figure 1c, the shutter either blocks ($x_i^\mu=0$) or allows ($x_i^\mu=1$) the probe beam to reach the CCD camera behind the selected sample area for the synaptic weight. These weights and the threshold are defined by illuminating different areas of the sample with the pump pulses using a translation stage. For convenience, the threshold was stored after initialization. A physical separation of 136 μm between the weights was chosen to minimize the influence of scattered light from the shutter edge.

The sum of the integrated intensities can in principle be directly evaluated using a photo-diode. Here instead we directly work with integrated intensities from the CCD camera and evaluate the weighted sum and subtraction as well as the sign function using an external LabVIEW program. This allows us to directly extract the weight evolution as described further below. By comparing the output with the desired output, we obtain the error signal. Subsequently, weights are adjusted by sending pump pulses when the input and error are nonzero. The sign of the error is fed back by the computer-controlled orientation of the λ/4 plate, ensuring that weights are either increased or decreased. We adjusted the learning rate $\eta$ by the number of pump pulses. For our setup, pulse packets consisting of 5 pump pulses each of 4 ps duration and separated by 1 ms, yield reproducible results.

The goal of the learning scheme is to obtain the desired output for each of the input patterns by adapting the synaptic weights. This is demonstrated in Figure 3. Figures 3a(3d) and 3c(3f) show the magneto-optical images recorded before and after the learning of the AND (OR) function, respectively. In each of these panels, rows 1-3 show recorded images of weighted inputs $y_1$, $y_2$ and threshold -b, respectively. The columns indicate the different input patterns. For each pattern, the sum $y_1+y_2-b$ is shown in the fourth row. These are used to evaluate the normalized integrated pixel intensity signal shown in the bottom row. This integrated intensity is negative for all input patterns prior learning, as seen in Figure 3a and 3d.

In order to learn the AND function, weights are updated according to the perceptron learning rule (Eq. (2)), which uses 1 pulse packet of 5 pulses as the learning rate. Figure 3b shows the evolution of the integrated intensity of the measured weighted sum $y_1^\mu + y_2^\mu - b$ as a function of the number of learning steps. For each step, only one pattern µ is evaluated and subsequent steps are obtained by sequentially cycling through the input patterns. After 16 learning steps the AND function is learned. This is directly verified from the integrated pixel intensities shown in the bottom of Figure 3c, displaying a positive sign of the integrated pixel intensity only when both inputs are one (µ=4). Similarly, an OR gate was learned, for which 70 learning steps were needed (Figure 3e). Again, this is directly verified from the bottom panel of Figure 3f, showing positive integrated intensities except when both inputs are zero (µ=1).

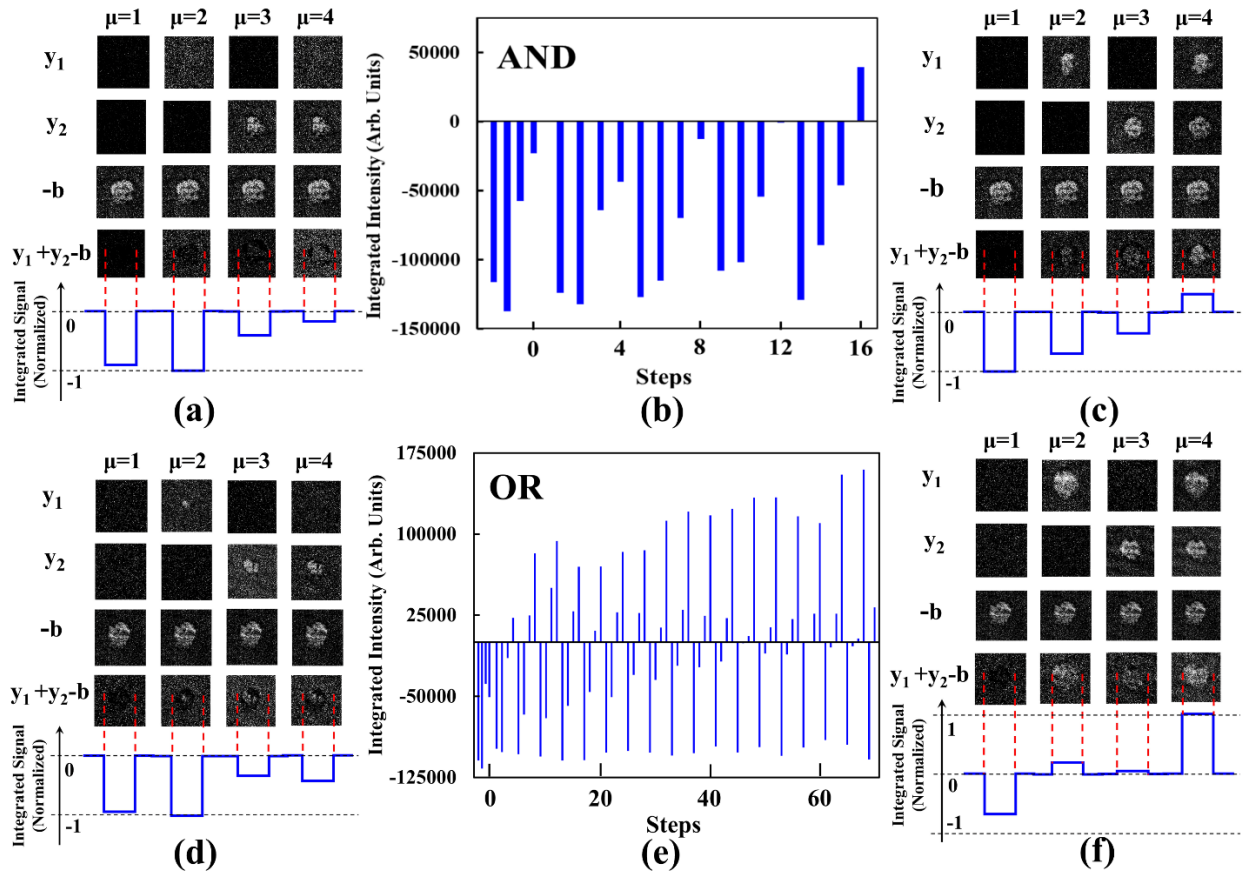

Figure 3: Supervised learning in the magneto-optical setup. (a) and (c) show the background subtracted CCD images before and after the learning of the AND function, respectively. The four figures from left to right denote the input patterns (0, 0), (1, 0), (0, 1), and (1, 1). Weighted inputs $y_1$ and $y_2$ for these patterns are shown in the first and second

row, respectively, while the third shows the threshold b. The fourth row shows the weighted sum $y_1+y_2-b$. The bottom row shows the normalized integrated intensities calculated from the fourth row. Panel (b) shows the evolution of the integrated intensity $y_1+y_2-b$ during learning, which proceeds by cycling through the input patterns. (d-f) show the corresponding results for the implementation of the OR function.

To infer how accurately the experimental implementation follows the perceptron learning rule we compared the evolution of weights seen experimentally with those from numerical simulation. The latter is a software implementation in MATLAB of the supervised learning algorithm described above with the learning rules Eqs.(1)-(2). To facilitate the comparison, we normalized the integrated intensity experimentally determined in Figure 2. For the AND gate, we obtained ($w_1$, $w_2$) = (0.0027, 0.27) and the threshold b=0.66. Using a learning rate of η=0.0665, the simulation learns the AND function with the same number of learning steps as in the experiment. The evolution of the weights is plotted in Figure 4a, which shows experimental and simulated weights as red triangles and black stars, respectively. Experimental error bars take into account intensity fluctuations in the background subtracted CCD images as measured by blocking the pump beam. We observe that unlike the simulated perceptron, in practice the weights do not exhibit a strictly monotonic increase during learning. Deviations from strictly linear weight changes are seen also in Figure 2. They can be modeled as variations in the learning rate due to changes in the pump polarization, but do not give rise to non-monotonic weight variations. Instead, we attribute this to light scattering from the shutter, which is present for the patterns $\mu = 2,3$ but not for $\mu = 4$ (note that for $\mu = 1$ no weight measurements take place since inputs are blocked by the shutter), leading to smaller measured weight values for pattern $\mu = 4$. The influence of such scattering is stronger when more learning steps are needed as we explored for the OR gate in Figure 4b. In this case, the normalized threshold is b=0.58 and the normalized initial weights are the same as for the AND gate. Without accounting for light scattering, the simulations learn the OR function 36 steps faster than seen in experiment (data not shown). If, instead we account for a systematic offset in the weight measurement for μ=4: $w_i \rightarrow w_i + c_4$, where $c_4 = -0.027$, we recover qualitatively similar oscillatory weight changes during the learning process. As shown in Figure 4b, this also proceeds with the same total number of learning steps as in the experiment, supporting the conjecture that the scattered light is the source of the non-monotonic evolution of weights. Importantly, despite the presence of these experimental non-idealities that slow down the learning, convergence is still reached. This shows that the perceptron learning rule is rather robust against device imperfections, suggesting that upscaling to larger opto-magnetic neural networks might well be feasible.

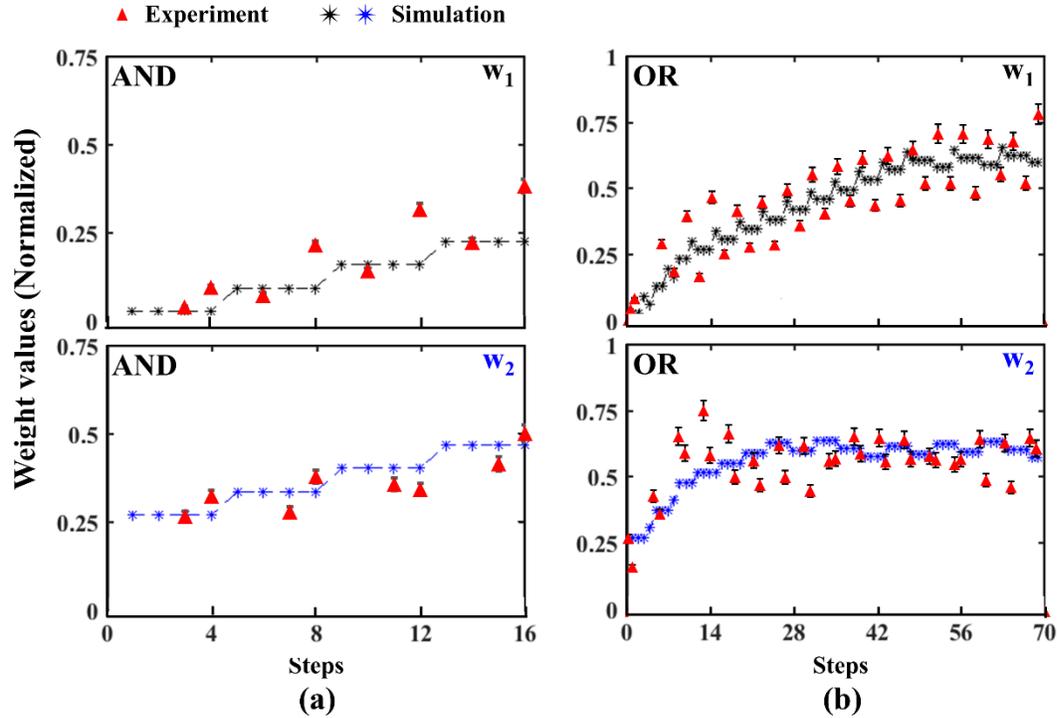

**Figure 4:** Evolution of synaptic weights during learning for (a) AND function and (b) OR function. Red triangles show experimental results $y_i = w_i x_i$, that are only detectable when $x_i = 1$ (small values $y_i$ at $x_i = 0$ are not included for clarity). Stars indicate the results from computer simulations.

In summary, we have demonstrated an opto-magnetic implementation of an artificial neural network. As an application, we demonstrated supervised learning of the Boolean AND and OR functions. The synaptic weights are stored non-volatile and adapted reversibly using ultrashort circularly polarized laser pulses exploiting a global feedback mechanism. These results suggest high potential for improving the energy-efficiency and learning speed of artificial neural networks. In the current setup, the energy absorption per learning step is estimated from the optical fluence and diameter used for the synaptic weights (1.125μm), yielding 64pJ/synapse/step. The time between successive learning steps is currently limited by the translation stages, being in the order of seconds. In principle, much faster control of light is feasible using wavefront shaping[19, 20]. Ultimately, the learning speed is only limited by the optical pulse duration, yielding rates as high as 50 GHz for the current setup. It will be very interesting to investigate downscaling the synaptic weight area to reduce the energy cost and time required per learning step, potentially down or even below the femtojoule and picosecond regime for nanoscale synapses.

**Acknowledgements**

We would like to acknowledge discussions with Bert Kappen, Misha Katsnelson, Alex Khajetoorians and Riccardo Zecchina, as well as F. Ando and T. Ono for sample growth. This research was in part supported by the ERC Advanced Grant 339813 (EXCHANGE), the Nederlandse Organisatie voor Wetenschappelijk Onderzoek (NWO) and the Nationale Wetenschaps Agenda (NWA, startimpuls GreenICT). J.H.M. acknowledges funding from NWO by a VENI grant and is part of the Shell-NWO/FOM-initiative "Computational sciences for energy research" of Shell and Chemical Sciences, Earth and Life Sciences, Physical Sciences, FOM and STW. Technical support from Tonnie Toonen, Chris Berkhout and Sergey Semin is gratefully acknowledged.